\documentclass[final,3p,times,twocolumn]{elsarticle}
\usepackage{amssymb}

\usepackage{tikz}

\usepackage{filecontents}

\usepackage{graphicx}
\usepackage[linesnumbered,ruled,vlined]{algorithm2e} %
\usepackage{booktabs,multirow} 
\usepackage{hyperref}  
\usepackage{mathrsfs}
\usepackage{float}
\usepackage{wrapfig}
\usepackage{adjustbox}
\usepackage{amsmath}

\journal{}

\begin{document}

\begin{frontmatter}



\title{A clean-label graph backdoor attack method in node classification task}


\author{Xiaogang Xing,  Ming Xu\textsuperscript{*}, Yujing Bai, Dongdong Yang}

\address{School of Cyberspace, Hangzhou Dianzi University, Hangzhou 310018, China}
\address{Army Engineering University of PLA Communication NCO Academy, Chongqing 400036, China}

\begin{abstract}
Backdoor attacks in the traditional graph neural networks (GNNs) field are easily detectable due to the dilemma of confusing labels.
To explore the backdoor vulnerability of GNNs and create a more stealthy backdoor attack method, a clean-label graph backdoor attack method(CGBA) in the node classification task is proposed in this paper.
Differently from existing backdoor attack methods, CGBA requires neither modification of node labels nor graph structure.
Specifically, to solve the problem of inconsistency between the contents and labels of the samples, CGBA selects poisoning samples in a specific target class and uses the label of sample as the target label (i.e., clean-label) after injecting triggers into the target samples. To guarantee the similarity of neighboring nodes, the raw features of the nodes are elaborately picked as triggers to further improve the concealment of the triggers.
Extensive experiments results show the effectiveness of our method.
When the poisoning rate is 0.04, CGBA can achieve an average attack success rate of 87.8\%, 98.9\%, 89.1\%, and 98.5\%, respectively.
\end{abstract}

\begin{keyword}
Machine learning \sep Network security \sep  Graph neural networks   \sep Backdoor  \sep Node classification



\end{keyword}

\end{frontmatter}



\section{Introduction}
%-------------------------------------------------------------------------------

Graph Neural Networks (GNNs) have been demonstrating powerful modeling and prediction capabilities in multiple domains such as social network analysis\cite{1kumar2022influence, 2zhang2022improving}, recommender systems\cite{3gao2022graph, 4wu2022graph}, and chemical molecule prediction  \cite{5hao2020asgn, 6wang2022molecular}. However, similar to other machine learning models, GNNs face many security and privacy challenges. Backdoor attacks have attracted much attention as one of these potential threats.

Backdoor attacks aim to produce misleading results for a model under specific input samples or conditions by purposefully tampering with the training data or the model. 
While traditional backdoor attacks focus on tasks such as image \cite{7feng2022fiba,8hua2023unambiguous,9gong2023kaleidoscope} and natural language processing\cite{10li2022backdoors,11pan2022hidden}, there is a scarcity of research on GNNs.
In graphs, the relationships between nodes and edges contain rich information, and GNNs extract useful feature representations by learning from these relationships. However, this property also exposes potential opportunities for attackers to launch backdoor attacks against GNNs by tampering with the graph structure or feature representations in a targeted manner. Once the backdoor is successfully embedded in the model, an attacker can mislead the GNNs to produce incorrect outputs by triggering specific input patterns, which can lead to serious security issues. For example, in social networks, an attacker can add fake users to the network. There are some hidden connections between these fake users and real users, which may lead to false positives or omissions in anomaly detection by the GNN model.

Although there have been some explorations on backdoor attacks on GNNs\cite{12xi2021graph,13zheng2023link,14dai2023semantic,15zhang2021backdoor,16zheng2023motif,17alrahis2023poisonedgnn,18chen2022neighboring,34zhang2023graph}, generally still in their infancy. Moreover, the majority of research focuses on graph classification, while backdoor attacks on node classification tasks are still very limited.
Chen et al\cite{19chen2023feature}. propose a backdoor attack method NFTA on node classification task using node features as triggers, which modifies the graph structure to avoid feature triggers from destroying the similarity between nodes.
Dai et al\cite{20dai2023unnoticeable} also believe that the insertion of triggers destroys the similarity between nodes, which will result in triggers that are easy to detect and unable to resist adversarial defense methods based on node similarity. 
To build an unnoticeable attack method, they propose UGBA (Unnoticeable Graph Backdoor Attack) from the perspective of limited attack budget and backdoor Unnoticeability.
Existing methods of backdoor attacks on node classification tasks need to modify the labels of the nodes. In addition, to guarantee the similarity of nodes, they either tweak the graph structure or require additional optimization of feature triggers, and such attacks also have higher graph perturbation and attack overhead.
To exploit the backdoor vulnerability of GNNs and build a simpler, more effective, and more stealthy backdoor attack, we propose a novel \underline{\textbf{c}}lean-label \underline{\textbf{g}}raph \underline{\textbf{b}}ackdoor \underline{\textbf{a}}ttack (CGBA) method for the node classification task.

Unlike existing node classification backdoor attacks, CGBA is a simple and effective attack method that requires neither modification of node labels nor additional optimization of feature triggers and graph structures. CGBA achieves efficient attack performance and invisibility by injecting only subtle triggers.

The difference between our approach and existing backdoor attacks is shown in Table \ref{tab1}.
\begin{table}[!htbp]
	\caption{ Similarities and differences between backdoor attacks on existing node classification tasks and our method} 
	\label{tab1}
	\centering
	
	\resizebox{0.4\textwidth}{!}{
		\begin{tabular}{cccc} 
		\toprule 
		& NFTA & UGBA  & Ours\\
		\midrule 
		Node label modification   &   Y   &   Y     &  N  \\
		Graph structure modification &   Y   &   Y     &  N \\
		Node similarity      &   Y   &   Y     &  Y  \\

		\bottomrule 
	\end{tabular}
	}
\end{table}

Specifically, we select poisoned nodes in a specific target class. Without modifying the node label after implanting a trigger on the poisoned node, we use the node's own label as the target label, i.e., the clean-label.
CGBA selects triggers among the raw features of nodes with larger degrees and does so for two reasons. First, according to the homogeneity assumption, nodes with the same label are usually more similar\cite{22li2022reliable}, while the insertion of feature triggers may change the similarity\cite{20dai2023unnoticeable}. Therefore, choosing raw features as triggers has less impact on the similarity between nodes, and also makes the presence of feature triggers more natural as well. Second, the importance of a node with a larger degree is higher and the node has more neighboring nodes (i.e., more similar nodes), thus using that node's features as a trigger will further enhance the attack's insidiousness.

The contribution of this paper is as follows:
\begin{itemize}
	\item We propose CGBA, a clean-label backdoor attack on the node classification task, thereby enhancing the stealthiness of the backdoor.
	
	\item CGBA guarantees the similarity of features among neighboring nodes and the effectiveness of the attack without optimizing the graph structure and feature triggers, simplifying the attack process.
	
	\item Extensive experiments on multiple benchmark datasets are conducted to validate the effectiveness of the CGBA, and it exhibits significant robustness when facing defense methods based on node similarity.
\end{itemize}

The rest of this paper is organized as follows: In Section \ref{sec2}, we briefly review related works. Then, in Section \ref{sec3}, we introduce the related preliminaries and definitions. Our proposed core methodology is described in detail in Section \ref{sec4}. To validate the effectiveness of our methodology, extensive experiments and analyses are conducted on multiple benchmark datasets and models in Section \ref{sec5}. Finally, we summarize the paper in Section \ref{sec6}.

\section{Related works}
\label{sec2}
Backdoor attacks can be categorized into two types according to whether the labels are modified or not. One category is the dirty-label backdoor attack, which refers to the modification of the sample label to the target class after the trigger is implanted into the sample; the other category is clean-label backdoor attacks, which means that the trigger is injected into a specific class of samples, and then the original label of the sample is utilized as the target label.

\subsection{Dirty-label backdoor attack in GNNs}

\indent \textbf{Graph Classification}: A GNN-oriented backdoor attack method is proposed by Zhang et al\cite{15zhang2021backdoor}, employing fixed subgraphs as triggers. However, the use of fixed triggers renders the attack susceptible to detection. In GTA\cite{12xi2021graph}, adaptive subgraphs are introduced as triggers, effectively enhancing the stealthiness of the attack. The impact of trigger injection at different locations on attack effectiveness is explored by Xu et al \cite{31xu2021explainability} from an interpretability perspective, with the aim of identifying optimal triggers. In contrast, Jiang et al\cite{32jiang2022defending} employ interpretability to identify potential trigger locations, thus defending against GNN backdoor attacks. Given that backdoor attacks necessitate the correlation of triggers with labels, the Graph Contrastive Learning defense method is proposed by Chen et al\cite{33chen2023contrast}, which does not rely on sample labels.

\noindent \textbf{Node classification}: Currently, there are fewer methods for backdoor attacks in node classification tasks. In NFTA\cite{19chen2023feature}, some of the features of the nodes are employed as triggers, and the node labels are modified to the target class after implanting the triggers. To guarantee that modifying labels does not destroy the similarity between nodes, the graph structure is adaptively adjusted in NFTA. Since the insertion of triggers destroys the similarity between nodes, Dai et al\cite{20dai2023unnoticeable} crop the edges between dissimilar nodes to effectively reduce the performance of the backdoor attack. To be able to create a more stealthy backdoor attack method, they propose the unnoticeable graph backdoor attack algorithm UGBA. UGBA introduces additional triggers to attach to the original graph. They optimize the triggers to ensure the similarity between nodes, which effectively improves the covertness.

\subsection{Clean-label backdoor attack}

\indent \textbf{Image Backdoor}: The modification of image labels in a backdoor attack results in a discrepancy between the actual image content and the assigned label, making manual detection relatively straightforward. To circumvent manual detection, the clean-label backdoor attack method is proposed by Turner\cite{26turner2018clean}, avoiding direct alteration of labels. In a bid to further bolster concealability, Ning et al\cite{35ning2021invisible} simulate triggers as imperceptible noise, thereby augmenting the stealthiness of the triggers. Nevertheless, clean-label attack methods typically struggle to achieve a high attack success rate, as the poisoned nodes are confined to the target class. Consequently, the model tends to prioritize the learning of more resilient original features over triggers. Addressing this limitation, Gao et al\cite{36gao2023not} advance the clean-label backdoor attack by strategically poisoning challenging samples instead of selecting them randomly.

\noindent \textbf{Graph backdoor}: At present, there is only one work for clean-label backdoor attacks in graph classification tasks \cite{25xu2022poster}.
This scheme used subgraphs as triggers and injected triggers in specific target classes.

To the best of our knowledge, clean-label backdoor attacks have not been explored on node classification tasks, which is one of the motivations for our work. To bridge this gap, CGBA is proposed in this paper.

%-------------------------------------------------------------------------------
\section{Preliminaries and problem formulation }
\label{sec3}
%-------------------------------------------------------------------------------
$\textbf{Notations}$. For a graph $ \mathcal{G = (V,E,X) }$ with $n$ nodes, where $\mathcal{V}$ represents the node set, $\mathcal{V}=\{v_1,v_2, ...,v_n\}$; $\mathcal{E}$ represents the edge set, $\mathcal{E} = \{e_1,e_2, ...,e_m\} $; $\mathcal{X} \in \mathbb{R}^{n*d}$ is the feature matrix consisting of node feature vectors, $\mathcal{X}=\{x_1, x_2, …, x_n \}^T$, where $x_i$ represents the feature vector of node $v_i$, with each vector have a dimension of $d$. The adjacency matrix $\mathcal{A} \in \mathbb{R}^{n*n}$ represents the structural information of the graph, where $\mathcal{A}_{ij}=1$ indicates the existence of an edge between nodes $i$ and $j$, while $\mathcal{A}_{ij} = 0$ signifies the absence of an edge between nodes $i$ and $j$. In GNNs, a graph $\mathcal{G}$ is taken as input, comprising the adjacency matrix $\mathcal{A}$ and node features $\mathcal{X}$. Subsequently, a low-dimensional representation (embedding) is generated for each node $v$ by GNN. Within the GNN framework, node representations are updated by aggregating information from their respective neighbors\cite{23kipf2016semi,24velivckovic2017graph}. The ultimate node representations are employed for specific downstream tasks.

\noindent $\textbf{Threat models}$. The threat model in this work considers a gray-box model where the attacker cannot access any knowledge of the relational model but accesses the training dataset. Given a graph, the attacker can inject triggers into the target sample. The aim of the attacker is to expect the classifier to classify the samples with triggers into target label, and classify normally for samples that do not carry triggers.

\noindent $\textbf{Dirty-label backdoor attack in node classification}$. In the backdoor attack for the node classification task, $g$ denotes the trigger, and $l(v)$ denotes the label of the node $v$. $\mathcal{M}(v, g)$ is a mix function that denotes the injection of the trigger $g$ into the feature vector of the node $v$. In a regular dirty-label backdoor attack, the label is modified to the target label $y_{t}$ after the trigger is injected into the sample, i.e.: 
\begin{align} 
	\label{eqution6}
	\begin{cases}
		l(v)=y_{truth}  \\
		l(\mathcal{M}(v,g))= y_{t}
	\end{cases}
\end{align}

In this work, the model $f_b$ obtained from the poisoning graph training is termed as the backdoor model and $f_c$ denotes the clean model. In the testing phase, unlabeled nodes with triggers are fed to the backdoor model $f_b$, which is misled and classifies the labels of the unlabeled nodes into the target class $y_{t}$, and behaves normally for the prediction of clean nodes. The process can be written as:
\begin{align} 
	\label{eqution2}
	\begin{cases}
		f_b(\mathcal{M}(v,g))=y_{t}  \\
		f_b(v) = f_c(v)
	\end{cases}
\end{align}

Unlike the dirty-label backdoor attack, the clean-label backdoor attack uses the node's own label as the target label. We detail our clean-label backdoor attack method in Section \ref{sec4}.

%-------------------------------------------------------------------------------
\section{The proposed method}
\label{sec4}
%-------------------------------------------------------------------------------

\subsection{CGBA Overall Framework}	
In traditional backdoor attacks, the altered labels are inconsistent with the real contents of the samples, resulting in a backdoor that is easy to detect. Hence clean-label backdoor attacks aim to establish an association between the trigger and the original label to enhance the stealthiness of the backdoor and thus evade detection. Differently from regular backdoor attacks, the poisoned nodes in a clean-label backdoor attack are selected in a specific target class. Once the sample is injected into the trigger, the original label of the sample is utilized as the target label, i.e.:
\begin{equation}
	\label{eqution3}
	\mathcal{L}(v) = \mathcal{L}(\mathcal{M}(v,g)) = y_{truth} =y_{t} 
\end{equation}

\begin{figure*}[h]

	\centering
	
	\includegraphics[width=0.9\linewidth]{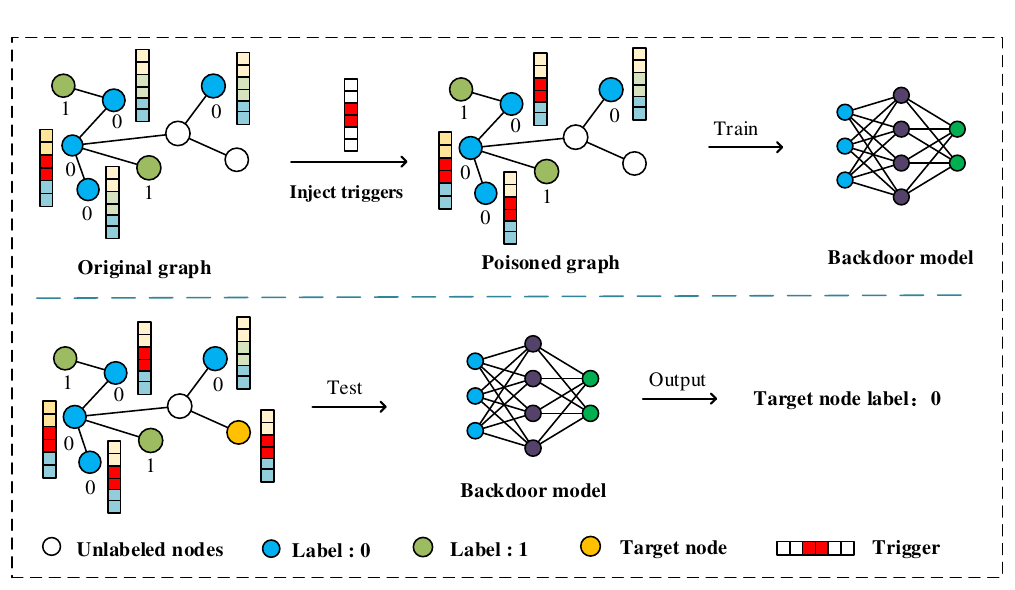}
	\caption{CGBA overall structure. The target node is the node that is injected with the trigger to test the attack effectiveness of the model. }
	\vspace{10pt} % adjust distance among fig to title
	\label{fig1}
	
\end{figure*}

The overall framework of the CGBA is shown in Figure \ref{fig1}. Feature triggers are injected in nodes with label 0 and then the graph is leveraged to train the backdoor GNN model. It is worth noting here that the node label is not modified after injecting the trigger and the label remains 0. When an unlabeled target node with a trigger is fed into the backdoor model for testing, the model classifies the target node with the target label 0.

The trigger size is usually small and the proportion of poisoning is relatively low. Hence without changing the label, the model is more inclined to ignore the trigger and associate the original features of the sample with the target label. To achieve a clean-label attack, the typical approach is to interfere with the original features so that the model cannot learn the original features correctly, forcing the model to correlate the trigger with the target label\cite{26turner2018clean}. However, we cannot significantly interfere with the original features in the graph data, which might change the feature similarity between nodes. Thus there is a major challenge in clean-label backdoor attacks in node classification: \textbf{How to make the model correlate triggers well with labels without perturbing the original features?}

To cope with this challenge, CGBA selects some of the robust features of nodes with larger degrees as triggers. First of all, the node's own robust feature usually has a strong association with its own label, and using this feature as a trigger is conducive for the model to establish an association between the trigger and the label.
Typically, features with larger values are more robust, so we choose the part of the original feature with the largest value as the trigger. Secondly, a node with a larger degree usually has more neighbors, which means that the node has more similar nodes.  Furthermore,  the node has a higher importance, and the training of the model is more influenced by this node.  Selecting the features of this node as a trigger is less influential on the node similarity. In addition, the original features are naturally present in the feature vector of the node, which contributes to enhancing the stealthiness of the trigger.

To enable a clearer presentation of CGBA, we formalize the trigger selection process. Assume that the target node is $v_t$ with a $d$-dimensional feature vector $x_t=(f_{t11}, f_{t12}, …, \mathbf{f_{t1i}}, …, f_{t1j}, … \mathbf{f_{t1k}},…, f_{t1d})$, and triggers are $g_t=(\mathbf{f_{t1i}}, \mathbf{f_{t1k}})$, $g_t \subset x_v$. The set consisting of $N$ poisoned nodes is $\mathcal{V}_P$ and the feature matrix is $\mathcal{X}_P$.

\begin{equation*}
	\mathcal{X}_P =
	\begin{bmatrix} 
		f_{t11} \;\; \cdots \;\; \mathbf{f_{t1i}} \;\; \cdots \;\; f_{t1j} \;\; \cdots \;\; \mathbf{f_{t1k}} \;\; \cdots  \;\; f_{t1d} \\
		f_{v21} \;\; \cdots \;\; f_{v2i}          \;\; \cdots \;\; f_{v2j} \;\; \cdots \;\; f_{v2k}          \;\; \cdots  \;\; f_{v2d} \\
		\vdots  \;\;\;\;\;\;\;\;\;\;\;\;\;\;\; \;\; \;\; \;\;\;\; \ddots \;\; \;\; \;\;\;\;\;\;\;\;\; \;\;\;\;\;\;\;\;\;\;\;\;\;\; \vdots   \\
		f_{vn1} \;\; \cdots \;\; f_{vni}          \;\; \cdots \;\; f_{vnj} \;\; \cdots \;\; f_{vnk}          \;\; \cdots  \;\; f_{vnd} \\
	\end{bmatrix}
\end{equation*}

Feature triggers in both the matrix and the vector are bolded.
The injection of a trigger can be formulated as:

\begin{equation*}
	\label{eqution4}
	\mathcal{M}(\mathcal{V}_P,g_t) = 
	\begin{bmatrix} 
		f_{t11} \; \cdots \; \mathbf{f_{t1i}}  \; \cdots \; f_{t1j} \; \cdots \; \mathbf{f_{t1k}} \; \cdots  \; f_{t1d} \\
		f_{v21} \; \cdots \; \mathbf{f_{t1i}}  \; \cdots \; f_{v2j} \; \cdots \; \mathbf{f_{t1k}} \; \cdots  \; f_{v2d} \\
	\vdots \;\;\;\;\;\;\;\;\; \;\; \;\; \;\;\;\; \ddots \;\; \;\; \;\;\;\;\;\;\;\;\; \;\;\;\;\;\;\;\;\;\;\;\vdots   \\
		f_{vn1} \; \cdots \; \mathbf{f_{t1i}}  \; \cdots \; f_{vnj} \; \cdots \; \mathbf{f_{t1k}} \; \cdots  \; f_{vnd} \\
	\end{bmatrix}
\end{equation*}

It is important to note that after the feature triggers are selected, they are not injected randomly, but at the positions in the feature vector corresponding to the feature triggers. This is since in the eigenvectors, the eigenvalues at each position usually represent similar meanings. It can be seen that the triggers chosen in CGBA can be discontinuous. For instance, there are $k-i$ intervals in the eigenvectors between the eigenvalues $f_{t1i}$ and $f_{t1k}$ in the trigger $g_t$. The purpose of this is to enhance stealth, as successive feature vectors are more conspicuous.

After the triggers are injected, the triggers are associated with the target labels by optimizing the cross-entropy loss. The loss can be formulated as:

\begin{equation}
	\label{eqution5}
	Min  \mathcal{L}(\theta) =    \sum_{v_i \in \mathcal{V}_P  } \mathcal{L} (f_b(\mathcal{M}(v,g_t)), y_{t}) 
\end{equation}

where $y_t$ denotes the target label, but in a clean-label backdoor attack, the target label is the true label of the node.

\subsection{Algorithm design}	
Algorithm \ref{algorithm1} describes the process of selecting and injecting triggers by CGBA. 

\begin{algorithm}[ht]  % Dont use [H] in two column
  \label{algorithm1}
	\caption{Clean-Label graph backdoor attack method in node classification task (CGBA)}
	\KwIn{A clean graph $ \mathcal{G_C = (V,E,X) }$, Target Class $\mathcal{T} $, $Trig \_ Size$, $ Num\_poison$}
	\KwOut{A backdoor graph $\mathcal{G_B}$}

	\For{idx in train\_indices}{
		\If{Node\_Label[idx] == $\mathcal{T}$  }{
			Search for the largest drgree node  $Max\_Node$
		}
	}
	nonzero \_indices = nonzero\_indices ($Max\_Node$)
	
	nonzero\_values = nonzero\_values ($Max\_Node$) 
	
	Sorted\_indices = Sort(nonzero\_values) 
	
	Select the largest feature of $Trig\_Size$ as trigger 
	
	$ //  \ Inject  \ trigger $ 
	
	poison\_intensity = 0
	
	\For{idx in train\_indices}{
		\If{Node\_Label[idx] == $\mathcal{T}$ and poison\_intensity $<$ Num\_poison}{
			
			Node\_X [idx, nonzero\_indices] = trigger
			
			poison\_intensity = poison\_intensity + 1	
		}
	}

\end{algorithm}

The algorithm first finds the node with the largest degree and obtains the non-zero eigenvalue and index of that node, since non-zero elements on the eigenvector are effective eigenvalues. The non-zero values are then sorted in numerical order of magnitude and the largest eigenvalue is selected as the trigger according to the trigger size set in advance.
Finally, we inject triggers into the target class to obtain the poisoned training dataset.
Note that the non-zero values of some nodes may be sparse. When the number of non-zero eigenvalues is less than the trigger size, we take randomly generated eigenvalues to make up the trigger size.
To guarantee trigger invisibility, the value of the randomly generated eigenvalues is limited to the value interval of the original non-zero eigenvalue.

\section{Experiments}
\label{sec5}

In this section, we conduct extensive experiments on four benchmark datasets to demonstrate the effectiveness of CGBA. Our evaluations mainly answer the following questions:

\noindent \textbf{R1}: How effective is the attack?

\noindent \textbf{R2}: How is the stealthiness?

\noindent \textbf{R3}: How does the clean-label backdoor attack method perform compared to the modified label backdoor attack method?

\noindent \textbf{R4}: How do hyperparameters affect the effectiveness of an attack?

\subsection{Experimental settings}
The main description here is the dataset, evaluation metrics, and parameters used in the experiments.

\noindent \textbf{Datasets}.
A total of four benchmark datasets are utilized in the experiments, Cora, CiteSeer, PubMed\cite{27yang2016revisiting}, and Flickr\cite{28zeng2019graphsaint}. Cora, CiteSeer, and PubMed are three citation networks, and Flickr is a large-scale graph that connects image captions with the same attributes. Detailed information on the four datasets is provided in Table \ref{tab2}.

\begin{table}[!htbp]
	\caption{Datasets} 
	\label{tab2}
	\centering
	\resizebox{0.4\textwidth}{!}{
	\begin{tabular}{ccccc} 
			
		\toprule 
		Datasets  &  Cora  &  CiteSeer  &  PubMed  &  Flickr \\
		\midrule 
		\# Nodes   &  2708  &  3327      &  19717  &  89250  \\
		\# Edges   &  5278  &  4552      &  44324  &  899756 \\        
		\# Classes &  7     &  6         &   3     &   7     \\
		\# Features&  1433  &  3703      &  500    &  500    \\
		\bottomrule
	\end{tabular}
	}
\end{table}

In a clean-label backdoor attack, the nodes that are poisoned belong to a specific class.  Table \ref{tab3}  shows the total number of nodes in the training set and the number of nodes in each class in the training set for each dataset.

The number of classes in the four datasets is 7, 6, 3, and 7, respectively, and "-" indicates that the class does not exist in the dataset. The number of nodes in each class in the Cora, CiteSeer, and PubMed is 20, and in the Flickr, there is a significant difference in the number of nodes in each class.

\begin{table}[!htbp]
	\caption{Number of nodes in each class} 
	\label{tab3}
	\centering
	\resizebox{0.4\textwidth}{!}{
	\begin{tabular}{ccccc} 
		\toprule 
		Datasets   &  Cora  &  CiteSeer  &  PubMed  &  Flickr \\
		\midrule 
		Num\_train &  140   &  120       &  60     &  44625    \\
		Class 0    &  20    &  20        &  20     &  2628     \\
		Class 1    &  20    &  20        &  20     &  4321     \\        
		Class 2    &  20    &  20        &  20     &  3164     \\
		Class 3    &  20    &  20        &  -      &  2431     \\
		Class 4    &  20    &  20        &  -      &  11525    \\
		Class 5    &  20    &  20        &  -      &  1742     \\
		Class 6    &  20    &  -         &  -      &  18814    \\
		\bottomrule 
	\end{tabular}
}
\end{table}

\noindent \textbf{Metrics}:
In our assessment, we primarily employ the following evaluation metrics: Attack success rate (ASR), Clean accuracy(CA), Clean accuracy drop (CAD), Attack accuracy after defense (DAA), and the discrepancy in model prediction accuracy for clean samples before and after defense (DPA).

ASR denotes the probability that the backdoor model $f_b$ predicts a sample with a trigger as the target class; CA denotes the prediction accuracy of the $f_b$ for clean samples; CAD denotes the difference between the prediction accuracy of the clean model for clean samples and the prediction accuracy of the $f_b$ for clean samples.
Backdoor attacks should satisfy a low CAD since the implantation of triggers should not affect the model's prediction results for clean samples as much as possible. 
If the injection of triggers causes the model's prediction accuracy on clean samples to be too low, the model could not be deployed, which in turn leads to the failure of the attack.
ASR can be formulated as:

\begin{equation}
	\label{eqution5}
	ASR =  \frac{\sum_{i=1}^n \mathbb{I} (f_b(v_i)=y_{t}) }{n} 
\end{equation}

Where $\mathbb{I}$ denotes the indicator function, and if * is True, then $\mathbb{I(*)}$ values 1. CAD can be formulated as:

\begin{equation}
	\label{eqution6}
	CAD = Acc_{f_c}(Clean) - Acc_{f_b} (Clean)
\end{equation}

Where $Acc_{f_c}(Clean)$ and $Acc_{f_b}(Clean)$ denote the prediction accuracy of the clean model $f_c$ and the backdoor model $f_b$ for clean samples, respectively.

DAA represents the attack success rate on models after the defense, with the defense method used in this paper being an adversarial defense approach based on pruning low-similarity edges. DPA signifies the extent of reduction in the model's predictive accuracy on clean data after undergoing data pruning.

The DPA can be written as:
\begin{equation}
	\label{eqution6}
	CAD = Acc_{f_b}(Clean) - DAcc_{f_c} (Clean)
\end{equation}

where $DAcc_{f_b} (Clean)$ denotes the prediction accuracy of the trained model on clean samples after defense cropping of the poisoned datasets.
A good defense approach requires not only a small DAA but also a small DPA. if the DPA is too high, it indicates that defense cropping has caused the model's accuracy on clean samples to drop too much.

\noindent \textbf{Parameters}: 
Three state-of-the-art models are used in our experiments, which are GraphSage\cite{29hamilton2017inductive}, ChebNet\cite{37Chebnet2016}, and ARMAConv\cite{38ARMA2021}. Other parameters are shown in Table \ref{tab4}.

\begin{table}[ht]
	\caption{Parameters} 
	\label{tab4}
	\centering
		\resizebox{0.35\textwidth}{!}{
	\begin{tabular}{cc} 
		\toprule 
		Type            &     Value        \\
		\midrule 
		Loss function   &  Cross entropy   \\
		Activation function  &      Relu      \\    
		Optimizer        &    Adam \cite{30kingma2014adam}     \\
		Epoch            &      400        \\
		Learning rate   &      0.01    \\
		Random seed     &       0   \\
		Similarity threshold T  &    0.2     \\
		Dropout          &    0.2    \\
		\bottomrule 
	\end{tabular}}
\end{table} 

\noindent \textbf{Baseline}:
At present, backdoor attack methods on node classification tasks are extremely limited, and there is not even any work on clean-label backdoor attacks.
Therefore, we extend the State-of-the-art attack method UGBA\cite{20dai2023unnoticeable}(WWW'23) to a clean-label attack as a baseline, which is called UGBA-C in this paper.The purpose of this is to make the comparison more fair. 
In NFTA\cite{19chen2023feature}(IJIS'23), the position of the trigger is randomly generated, so we also add CGBA-R, a variant of CGBA, as a baseline. CGBA-R uses the same triggers as the CGBA attack method, but the injection position of the feature triggers is randomized.

\subsection{Experiment results}

\subsubsection*{\noindent \textbf{R1: How effective is the attack?}}

To demonstrate the effectiveness of our method, we conduct extensive evaluations. The experiments contain a total of three models and four benchmark datasets, and the overall experimental results are shown in Table \ref{tab5}. PR(poisoning rate) is 0.04, and the trigger size is 0.05. The experimental results prove that our method achieves a favorable success rate of attack.
For example, in ChebNet, CGBA achieves an average attack success rate of 87.8\%, 98.9\%, 89.1\%, and 98.5\% on the four datasets, where the average attack success rate refers to the average success rate of the attacks on each class.
Obviously, our method outperforms CGBA-R in all cases, while for UGBA-C, the difference in the effectiveness of our attack on Cora, PubMed, and Flickr is not significant, but our attack on CiteSeer is significantly better than  UGBA-C.

Regarding the Cora and PubMed datasets and the models GraphSAGE and ChebNet, the prediction accuracy of UGBA-C for clean samples is slightly higher than that of our scheme, but in Flickr, the accuracy of our scheme for clean samples is higher than that of UGBA-C. There is little difference in the experimental results for other datasets and models.

\begin{table*}[!ht]
	\caption{Overall Attack Effectiveness. ASR denotes the attack success rate. CA denotes the prediction accuracy of the backdoor model on clean samples. PR(poisoning rate) is 0.04, and trigger size is 0.05.} 
	\label{tab5}
	\centering
	\begin{adjustbox}{width=0.95\textwidth}  % Adjust table size
		\begin{tabular}{c c c c c c c c } 
			\toprule 
			\multirow{2}*{Datasets}  &	\multirow{2}*{Class}  & \multicolumn{3}{c}{ASR(UGBA-C $|$ CGBA-R $|$ Ours)(\%)} & \multicolumn{3}{c}{CA (UGBA-C $|$ CGBA-R $|$ Ours)(\%)}   \\
			
			\cmidrule(lr){3-5} \cmidrule(lr){6-8}
			
	&         & GraphSAGE           &ChebNet                &ARMAConv               & GraphSAGE            &ChebNet           &ARMAConv       \\
			\midrule 
			\multirow{7}*{Cora}  
	&	0     & 87.1 $|$ 64.2 $|$ 95.0  & 91.3 $|$ 54.8 $|$ 97.5    & 92.3 $|$ 55.7 $|$ 98.0     &81.6 $|$ 72.4 $|$ 71.8   &80.6 $|$ 69.8 $|$ 76.6 & 70.2 $|$ 68.8 $|$ 72.5  \\
	&	1     & 90.7 $|$ 30.8 $|$ 92.8  & 93.7 $|$ 30.9 $|$ 94.9    & 94.1 $|$ 33.6 $|$ 94.6     &80.7 $|$ 72.8 $|$ 72.5   &80.7 $|$ 69.7 $|$ 69.4 & 72.5 $|$ 68.9 $|$ 73.6  \\
	&	2     & 70.1 $|$ 20.8 $|$ 47.6  & 82.2 $|$ 13.1 $|$ 70.7    & 88.3 $|$ 14.4 $|$ 74.2     &79.4 $|$ 72.9 $|$ 71.7   &78.5 $|$ 69.5 $|$ 68.9 & 69.8 $|$ 71.3 $|$ 70.4  \\
	&	3     & 91.2 $|$ 65.4 $|$ 87.8  & 92.9 $|$ 62.6 $|$ 93.8    & 96.1 $|$ 64.6 $|$ 93.9     &78.2 $|$ 72.6 $|$ 71.6   &79.6 $|$ 67.6 $|$ 65.9 & 65.4 $|$ 65.0 $|$ 65.2  \\
	&	4     & 81.0 $|$ 35.6 $|$ 84.3  & 89.9 $|$ 27.4 $|$ 85.7    & 90.9 $|$ 26.5 $|$ 88.8     &77.1 $|$ 73.0 $|$ 73.2   &78.6 $|$ 71.3 $|$ 71.7 & 66.7 $|$ 70.8 $|$ 70.2  \\
	&	5     & 83.4 $|$ 27.8 $|$ 79.2  & 92.3 $|$ 30.3 $|$ 87.5    & 88.4 $|$ 36.2 $|$ 89.5     &76.4 $|$ 72.3 $|$ 71.4   &78.7 $|$ 69.7 $|$ 69.1 & 72.3 $|$ 69.6 $|$ 65.7  \\
	&	6     & 69.8 $|$ 34.7 $|$ 76.2  & 80.9 $|$ 24.9 $|$ 84.8    & 82.0 $|$ 19.9 $|$ 86.3     &80.7 $|$ 72.6 $|$ 70.9   &81.9 $|$ 68.8 $|$ 68.1 & 74.7 $|$ 70.3 $|$ 67.2  \\
			\midrule 
			\multirow{6}*{CiteSeer}  
	&	0    &71.7 $|$ 57.7 $|$ 98.2  &74.8 $|$ 46.7 $|$ 98.8   & 75.8 $|$ 51.4 $|$ 99.2     & 72.2 $|$ 73.2 $|$ 72.5    & 73.0 $|$ 76.9 $|$ 70.6   & 67.7 $|$ 72.5 $|$ 70.2 \\
	&	1    &92.0 $|$ 72.5 $|$ 99.7  &95.7 $|$ 84.5 $|$ 99.8   & 94.9 $|$ 78.7 $|$ 99.8     & 70.6 $|$ 70.8 $|$ 70.4    & 71.3 $|$ 70.0 $|$ 68.3   & 68.0 $|$ 69.7 $|$ 69.0 \\
	&	2    &46.6 $|$ 37.8 $|$ 99.3  &45.3 $|$ 49.6 $|$ 98.8   & 45.2 $|$ 53.1 $|$ 99.2     & 71.0 $|$ 71.4 $|$ 71.6    & 72.2 $|$ 69.3 $|$ 69.6   & 68.7 $|$ 68.6 $|$ 69.4 \\
	&	3    &41.1 $|$ 53.8 $|$ 99.9  &39.6 $|$ 28.4 $|$ 99.8   & 45.2 $|$ 30.2 $|$ 99.9     & 70.9 $|$ 71.6 $|$ 71.9    & 71.9 $|$ 69.8 $|$ 68.2   & 66.3 $|$ 69.6 $|$ 67.8 \\
	&	4    &31.1 $|$ 51.9 $|$ 94.8  &33.0 $|$ 43.3 $|$ 98.6   & 29.9 $|$ 35.5 $|$ 98.9     & 73.7 $|$ 71.6 $|$ 71.3    & 73.9 $|$ 69.8 $|$ 68.5   & 68.0 $|$ 68.5 $|$ 68.7 \\
	&	5    &33.6 $|$ 52.3 $|$ 96.4  &34.1 $|$ 53.2 $|$ 97.8   & 35.2 $|$ 55.0 $|$ 98.0     & 72.2 $|$ 72.3 $|$ 71.1    & 73.5 $|$ 70.2 $|$ 70.6   & 71.9 $|$ 70.1 $|$ 69.5 \\
			
			\midrule 
			\multirow{3}*{PubMed}  
	& 0   &82.3 $|$ 68.3 $|$ 69.3  &89.8 $|$ 68.8 $|$ 91.0  &91.7 $|$ 75.0 $|$ 89.8          & 85.7 $|$ 76.4 $|$ 76.8   &86.9 $|$ 76.0 $|$ 77.0   &86.5 $|$ 77.1 $|$ 76.7  \\
	& 1   &84.9 $|$ 74.6 $|$ 85.2  &92.7 $|$ 69.8 $|$ 87.6  &94.7 $|$ 68.7 $|$ 87.6          & 86.3 $|$ 77.2 $|$ 76.2   &86.9 $|$ 75.8 $|$ 76.7   &79.2 $|$ 76.7 $|$ 75.6  \\
	& 2   &90.5 $|$ 59.2 $|$ 73.6  &94.8 $|$ 79.1 $|$ 88.8  &96.4 $|$ 76.8 $|$ 90.5          & 86.1 $|$ 76.3 $|$ 77.3   &87.1 $|$ 75.4 $|$ 75.4   &81.7 $|$ 76.2 $|$ 75.6  \\
			
			\midrule 
			\multirow{7}*{Flickr}  
	& 0   & 99.8 $|$ 0.0 $|$ 100.0   & 100.0 $|$ 0.0 $|$ 99.7   & 99.9 $|$ 0.0 $|$ 98.5     & 45.1 $|$ 49.6 $|$ 49.2   & 43.9 $|$ 48.3 $|$ 48.5  & 43.7 $|$ 45.3 $|$ 45.1 \\
	& 1   & 99.9 $|$ 0.0 $|$ 92.2    & 100.0 $|$ 0.1 $|$ 91.1   & 100.0 $|$ 0.0 $|$ 1.5     & 45.6 $|$ 49.1 $|$ 49.4   & 44.4 $|$ 48.4 $|$ 48.6  & 42.3 $|$ 45.0 $|$ 45.2 \\
	& 2   & 99.9 $|$ 0.0 $|$ 100.0   & 100.0 $|$ 0.3 $|$ 99.9   & 100.0 $|$ 0.0 $|$ 98.1    & 45.5 $|$ 49.0 $|$ 49.1   & 44.3 $|$ 48.6 $|$ 48.5  & 42.3 $|$ 45.4 $|$ 45.2 \\
	& 3   & 99.8 $|$ 48.3 $|$ 100.0  & 100.0 $|$ 47.2 $|$ 99.9  & 99.9 $|$ 0.0 $|$ 99.8     & 44.9 $|$ 48.5 $|$ 48.4   & 44.2 $|$ 47.2 $|$ 46.5  & 43.8 $|$ 47.8 $|$ 44.6 \\
	& 4   & 99.9 $|$ 99.7 $|$ 100.0  & 99.9 $|$ 99.2 $|$ 100.0  & 99.9 $|$ 98.9 $|$ 100.0   & 46.7 $|$ 48.3 $|$ 44.7   & 45.2 $|$ 46.7 $|$ 44.4  & 43.6 $|$ 44.4 $|$ 43.7  \\
	& 5   & 99.9 $|$ 6.0 $|$ 99.5    & 99.9 $|$ 21.8 $|$ 99.5   & 99.9 $|$ 0.0 $|$ 98.9     & 45.6 $|$ 48.5 $|$ 48.6   & 44.8 $|$ 46.3 $|$ 46.0  & 43.9 $|$ 44.6 $|$ 44.3 \\
	& 6   & 100.0 $|$ 82.5 $|$ 99.8  & 100.0 $|$ 80.5 $|$ 99.2  & 100.0 $|$ 84.8 $|$ 97.8   & 44.6 $|$ 49.0 $|$ 49.0   & 43.8 $|$ 47.9 $|$ 47.5  & 41.7 $|$ 45.3 $|$ 45.4 \\
			
			\bottomrule  
		\end{tabular}
	\end{adjustbox}
\end{table*}

Overall, our scheme can achieve a satisfactory attack success rate. Furthermore, our scheme slightly outperforms the state-of-the-art method UGBA-C and significantly outperforms CGBA-R.

\subsubsection*{\noindent \textbf{R2: How is the stealthiness?}}

There is no common metric for evaluating the stealthiness of a backdoor, and to demonstrate that the stealthiness of our method outperforms existing attack methods\cite{19chen2023feature,20dai2023unnoticeable}, we analyze it from the perspectives of label modification, data perturbation, and adversarial defense.

\noindent \textbf{Label modification}: Since the clean-label backdoor attack method does not modify the real label of the samples, this method does not have the problem of inconsistency between the label and the contents of the samples. Therefore, clean-label backdoor attacks are generally considered to be more stealthy than backdoor attack methods that modify labels. 

\noindent \textbf{Data Perturbation}: Existing methods require modification of the graph structure. Chen et al \cite{19chen2023feature} modified the graph structure to adapt to the similarity between nodes. In UGBA\cite{20dai2023unnoticeable}, triggers are constituted by extra injected nodes, this attack will result in a modified graph that does not have the same number of nodes as the original graph.
In our approach, only subtle feature triggers need to be injected without modifying the graph structure. In addition, the triggers we inject come from the node's own features rather than externally inserted triggers, and the feature values of the triggers are strictly limited to the range of the node's own features. This further enhances the invisibility of the trigger.

\noindent \textbf{Adversarial defense}: Since the insertion of feature triggers reduces the similarity between nodes, it makes the triggers susceptible to cropping by similarity-based adversarial defense methods\cite{20dai2023unnoticeable}. To demonstrate the strong trigger concealment of our method, defense experiments are conducted on four poisoned datasets using a defense method based on low-similarity edge pruning.
As shown in Table \ref{tab6} are the results of the defense experiments on the four datasets. From the experimental results, it can be learned that after defense cropping, the attack success rate not only does not decrease but also improves to some extent. This phenomenon is more obvious in Cora and CiteSeer.
This phenomenon occurs because the edges that are cropped out are not toxic, but clean edges in the graph. Therefore, instead of mitigating the toxicity, defense cropping amplifies the role of the trigger in the graph. In other words, the triggers in our method are more stealthy, and the cropping method based on low-similarity edges cannot effectively defend against the propagation of the triggers in the graph.

\begin{table}[!h]
	\caption{ Attack success rate before and after defense. ASR is the attack success rate and DAA is the attack success rate after defense pruning. Trigger size in the four datasets is 0.01, 0.005, 0.05, and 0.02, respectively. Pruning threshold T = 0.2 and poisoning rate PR = 0.06.  "-" indicates that the class does not exist in the dataset.} 
	\label{tab6}
	\centering
	\begin{adjustbox}{width=0.48\textwidth}  % Adjust table size
		\begin{tabular}{c c c c c c } 
			\toprule 
			\multirow{2}*{Model}  &	\multirow{2}*{Class}  & \multicolumn{4}{c}{ASR(\%) $|$ DAA(\%)} \\
			\cmidrule(lr){3-6}
			
			&	     & Cora         &CiteSeer       &PubMed        &Flickr                \\
			\midrule 
			\multirow{7}*{GraphSAGE}  
			&	0    &79.9 $|$ 91.5   &87.5 $|$ 96.2  &71.6 $|$ 82.9    &99.9 $|$ 99.9  \\
			&	1    &79.3 $|$ 90.9   &73.5 $|$ 93.8  &86.8 $|$ 87.2    &97.6 $|$ 96.7   \\
			&	2    &58.5 $|$ 91.5   &83.8 $|$ 95.0  &75.0 $|$ 89.2    &99.9 $|$ 99.9   \\
			&	3    &76.0 $|$ 91.6   &85.9 $|$ 97.9  &     -           &99.8 $|$ 99.8   \\
			&	4    &66.1 $|$ 81.6   &67.3 $|$ 92.7  &     -           &99.9 $|$ 99.9   \\
			&	5    &63.8 $|$ 83.3   &72.4 $|$ 95.9  &     -           &98.5 $|$ 99.6   \\
			&	6    &58.5 $|$ 81.6   &     -         &     -           &98.2 $|$ 97.6   \\
			\midrule 
			\multirow{7}*{ChebNet}  
			&	0     &86.4 $|$ 93.4  &69.1 $|$ 95.0   &89.7 $|$ 86.3    &99.8 $|$ 99.9   \\
			&	1     &80.5 $|$ 91.3  &86.9 $|$ 94.6   &89.0 $|$ 88.4    &96.5 $|$ 96.6   \\
			&	2     &70.6 $|$ 89.5  &85.4 $|$ 95.1   &91.5 $|$ 93.2    &99.8 $|$ 99.9   \\
			&	3     &79.3 $|$ 91.6  &90.2 $|$ 97.6   &     -           &99.8 $|$ 99.9   \\
			&	4     &64.5 $|$ 80.5  &73.7 $|$ 94.1   &     -           &99.9 $|$ 99.9   \\
			&	5     &68.1 $|$ 81.9  &81.0 $|$ 96.1   &     -           &98.8 $|$ 99.4   \\
			&	6     &64.1 $|$ 81.0  &     -          &     -           &97.1 $|$ 97.3   \\
			\midrule 
			\multirow{7}*{ARMAConv}  
			&	0     &85.8 $|$ 93.8  &89.6 $|$ 94.5   &89.6 $|$ 86.4    &90.8 $|$ 99.1   \\
			&	1     &80.3 $|$ 88.9  &84.3 $|$ 93.2   &88.9 $|$ 90.0    &34.0 $|$ 15.8  \\
			&	2     &73.8 $|$ 89.6  &86.1 $|$ 95.0   &90.1 $|$ 92.9    &99.8 $|$ 99.9   \\
			&	3     &80.4 $|$ 90.8  &89.8 $|$ 98.0   &     -           &99.8 $|$ 99.9   \\
			&	4     &70.2 $|$ 83.5  &77.0 $|$ 94.1   &     -           &99.9 $|$ 99.9   \\
			&	5     &71.6 $|$ 82.7  &83.5 $|$ 96.6   &     -           &98.6 $|$ 48.7   \\
			&	6     &66.8 $|$ 82.4  &     -          &     -           &95.5 $|$ 95.6   \\
			\bottomrule  
			
		\end{tabular}
	\end{adjustbox}
\end{table}

All of the above experiments are conducted under a fixed tailoring threshold, and to explore whether different pruning thresholds will affect the defense effect, we conduct experiments under different pruning thresholds.

\begin{figure*}[!ht]
	
	\centering
	\begin{minipage}[t]{0.9\linewidth} % 
		\centering
		\includegraphics[width=0.9\linewidth]{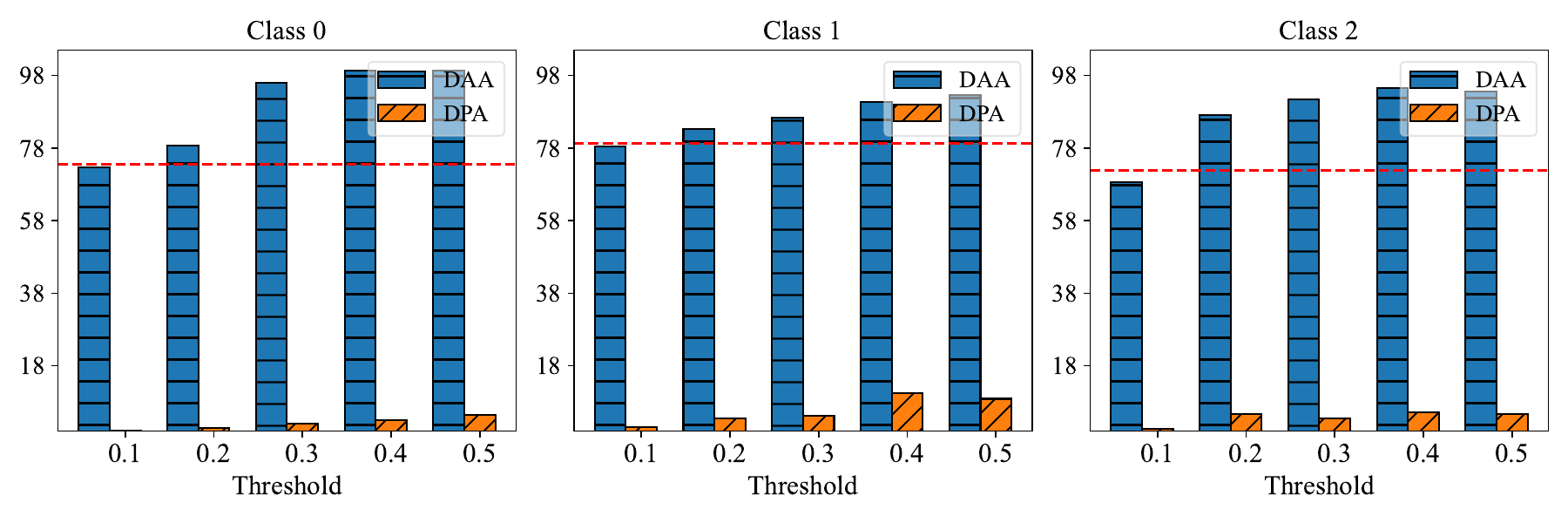}
		\label{fig2a}
	\end{minipage}
	\caption{The impact of different cropping thresholds on defense results in PubMed. The model is GraphSAGE. The poisoning rate is 0.04. Trigger sizes in four datasets are 0.01, 0.005, 0.05, and 0.02, respectively. DPA represents the difference in prediction accuracy of the backdoor model on clean samples before and after the defense.}
	\label{fig2}
\end{figure*}

As shown in Figure \ref{fig2} for the experimental results in  PubMed, the similarity thresholds are set from 0.1 to 0.5, respectively.
As for the training set, any edge below the threshold is cropped out. 
The red horizontal dashed line in each subfigure represents the attack success rate before the defense, DAA represents the attack success rate after the defense, and DPA represents the difference in prediction accuracy of the backdoor model on clean samples before and after the defense.
It can be seen that the attack success rate after cropping is still very close to the attack success rate before cropping when the threshold is 0.1. As the threshold value increases, the attack success rate not only does not decrease but also increases. The reason is that the CGBA backdoor attack method does not reduce the similarity between nodes, and the edges removed by defense cropping are clean in the original graph, not toxic edges.
Cropping causes the role of triggers in the original training set to be amplified, so there is a situation where the success rate of the attack is higher after cropping.
It can be observed that as the pruning threshold increases, DPA also increases. This indicates that as more clean edges are pruned, the reduction in predictive accuracy of the backdoor model for clean samples becomes more pronounced.

The results of the experiments on Cora, CiteSeer, and Flickr are shown in Table \ref{tab7}.  From the Cora, similar results to PubMed can still be obtained. In Flickr, the attack success rate of the model is not significantly increased after cropping since the model has a high attack success rate even before data cropping. But cropping also failed to make the accuracy rate decrease significantly, which means that the defense cropping method is still ineffective.

\begin{table*}[!ht]
	\caption{The effect of different pruning thresholds on defense effectiveness (DAA(\%) $|$ DPA(\%)). The model is GraphSAGE. The poisoning rate is 0.04. Trigger sizes are 0.01, 0.005, 0.05, and 0.02, respectively.} 
	\label{tab7}
	\centering
	\begin{adjustbox}{width=0.68\textwidth}  % Adjust table size
		\begin{tabular}{c c c c c c c c } 
			\toprule 
			\multirow{2}*{Datasets}  &	\multirow{2}*{Class} &	\multirow{2}*{ASR}  & \multicolumn{5}{c}{Threshold} \\
			\cmidrule(lr){4-8}
			
			&	&	& 0.1         &0.2        &0.3         &0.4         &0.5           \\
			\midrule 
			\multirow{7}*{Cora}  
			&	0  & 74.3   &77.2 $|$ 1.8 &88.3 $|$ 5.0 &94.5 $|$ 7.9 &95.5 $|$ 13.3 &96.2 $|$ 15.9   \\
			&	1  & 85.8   &87.2 $|$ 2.2 &87.8 $|$ 3.9 &93.2 $|$ 9.6 &94.5 $|$ 12.8 &95.0 $|$ 15.3  \\
			&	2  & 64.6   &67.8 $|$ 2.0 &89.2 $|$ 4.6 &94.4 $|$ 9.1 &95.4 $|$ 13.2 &95.2 $|$ 16.5   \\
			&	3  & 78.1   &76.3 $|$ 2.8 &88.9 $|$ 4.4 &92.5 $|$ 9.3 &95.4 $|$ 11.7 &94.4 $|$ 15.3   \\
			&	4  & 65.1   &64.9 $|$ 1.9 &73.4 $|$ 5.8 &85.5 $|$ 9.8 &87.0 $|$ 13.8 &92.7 $|$ 17.2   \\
			&	5  & 59.2   &71.3 $|$ 1.5 &82.5 $|$ 5.6 &91.9 $|$ 8.4 &94.3 $|$ 13.9 &94.9 $|$ 16.7   \\
			&	6  & 57.3   &60.7 $|$ 1.7 &79.9 $|$ 4.0 &87.8 $|$ 9.6 &90.7 $|$ 13.4 &91.2 $|$ 16.9   \\
			
			\midrule 
			\multirow{6}*{CiteSeer}  
			&	0  &78.2   &76.4 $|$ 0.2  &92.3 $|$ 1.5   &96.1 $|$ 4.0    &99.3 $|$  8.7    &99.7 $|$  10.1  \\
			&	1  &60.8   &73.6 $|$ 2.5  &87.8 $|$ 2.0   &94.2 $|$ 7.1    &98.0 $|$  9.4    &96.6 $|$  11.3  \\
			&	2  &70.1   &75.3 $|$ 0.3  &90.5 $|$ 1.7   &93.8 $|$ 5.9    &98.6 $|$  8.7    &98.5 $|$  10.8  \\
			&	3  &83.0   &91.6 $|$ 0.6  &97.0 $|$ 3.9   &98.6 $|$ 5.3    &99.5 $|$  9.2    &99.8 $|$  11.6  \\
			&	4  &67.7   &78.7 $|$ 0.5  &86.5 $|$ 1.7   &96.6 $|$ 6.2    &98.1 $|$  9.4    &99.6 $|$  10.5  \\
			&	5  &71.0   &75.6 $|$ 1.3  &92.3 $|$ 1.4   &95.2 $|$ 8.5    &97.7 $|$  11.9   &98.1 $|$  12.4  \\

			\midrule 
			\multirow{7}*{Flickr}  
			&	0  & 97.5   &95.8 $|$ 0.2 &95.5 $|$ 0.8 &97.8 $|$ 1.6 &98.3 $|$ 1.9 &97.6 $|$ 1.7   \\
			&	1  & 38.3   &32.6 $|$ 0.7 &46.2 $|$ 0.1 &36.9 $|$ 2.1 &40.6 $|$ 2.5 &45.8 $|$ 1.6   \\
			&	2  & 99.8   &99.7 $|$ 0.6 &99.7 $|$ 0.5 &99.7 $|$ 2.0 &99.5 $|$ 2.9 &99.8 $|$ 2.4   \\
			&	3  & 99.6   &99.3 $|$ 0.7 &99.5 $|$ 0.7 &99.7 $|$ 0.9 &99.6 $|$ 1.8 &99.7 $|$ 1.7   \\
			&	4  & 100.0  &100.0 $|$ 0.1&99.9 $|$ -0.2&99.7 $|$ 0.3 &99.9 $|$ 0.5 &99.5 $|$ 0.1  \\
			&	5  & 98.6   &97.5 $|$ 0.0 &99.7 $|$ 1.3 &99.7 $|$ 0.2 &99.6 $|$ 0.4 &99.7 $|$ 0.7   \\
			&	6  & 96.8   &97.3 $|$ -0.3&97.2 $|$ 1.2 &96.0 $|$ 1.6 &96.6 $|$ 1.7 &96.7 $|$ 2.0   \\
			
			\bottomrule  
		\end{tabular}
	\end{adjustbox}
\end{table*}

Through extensive experiments on various datasets, it can be concluded that the adversarial defense method based on low similarity edge cropping cannot resist our backdoor attack method.

\subsubsection*{\noindent \textbf{R3: How does the clean-label backdoor attack method perform compared to the modified label backdoor attack method?}}

Based on the above analysis, we know that the clean-label backdoor attack is more covert. 
To investigate the performance gap between the clean-label backdoor attack and the dirty-label backdoor attack, we compare the attack effectiveness of the two attack methods in the same setting.
It is worth noting that the poisoned samples for the dirty-label backdoor attack are not selected among the target classes, but are chosen randomly. The experimental results are shown in Table \ref{tab8}.

\begin{table*}[!ht]
	\caption{The performance gap between the clean-label backdoor attack and the dirty-label backdoor attack. PR is 0.04 and Trigger size is 0.05.  "-" indicates that the class does not exist in the dataset.} 
	\label{tab8}
	\centering
	\begin{adjustbox}{width=0.6\textwidth}  % Adjust table size
		\begin{tabular}{c c c c c c } 
			\toprule 
			\multirow{2}*{Model}  &	\multirow{2}*{Class}  & \multicolumn{4}{c}{ASR(Clean-label $|$ Dirty-label)(\%)} \\
			\cmidrule(lr){3-6}
			
			&	     & Cora         &CiteSeer       &PubMed        &Flickr                \\
			\midrule 
			\multirow{7}*{GraphSAGE}  
			&	0    &95.0 $|$ 96.8   &98.2 $|$ 100.0  &69.3 $|$ 78.0    &100.0 $|$ 100.0  \\
			&	1    &92.8 $|$ 98.2   &99.7 $|$ 99.8   &85.2 $|$ 89.1    &92.2 $|$ 97.4  \\
			&	2    &47.6 $|$ 98.5   &99.3 $|$ 98.7   &73.6 $|$ 79.6    &100.0 $|$ 100.0   \\
			&	3    &87.8 $|$ 96.3   &99.9 $|$ 98.4   &     -           &100.0 $|$ 100.0   \\
			&	4    &84.3 $|$ 93.9   &94.8 $|$ 100.0  &     -           &100.0 $|$ 100.0   \\
			&	5    &79.2 $|$ 91.7   &96.4 $|$ 99.9   &     -           &99.5 $|$ 99.0   \\
			&	6    &76.2 $|$ 95.3   &     -          &     -           &99.8 $|$ 99.9  \\
			\midrule 
			\multirow{7}*{ChebNet}  
			&	0     &97.5 $|$ 97.0  &98.8 $|$ 99.9   &91.0 $|$ 92.9    &99.7 $|$ 99.9   \\
			&	1     &94.9 $|$ 98.1  &99.8 $|$ 99.0   &87.6 $|$ 92.8    &91.1 $|$ 97.7   \\
			&	2     &70.7 $|$ 98.8  &98.8 $|$ 99.2   &88.8 $|$ 93.3    &99.9 $|$ 100.0   \\
			&	3     &93.8 $|$ 96.6  &99.8 $|$ 98.7   &     -           &99.9 $|$ 100.0  \\
			&	4     &85.7 $|$ 96.7  &98.6 $|$ 100.0  &     -           &100.0 $|$ 100.0  \\
			&	5     &87.5 $|$ 94.2  &97.8 $|$ 99.9   &     -           &99.5 $|$ 99.4   \\
			&	6     &84.8 $|$ 96.1  &     -          &     -           &99.2 $|$ 99.1  \\
			\midrule 
			\multirow{7}*{ARMAConv}  
			&	0     &98.0 $|$ 98.4  &99.2 $|$ 100.0  &89.8 $|$ 92.7    &98.5 $|$ 52.1  \\
			&	1     &94.6 $|$ 98.4  &99.8 $|$ 99.4   &87.6 $|$ 93.3    &1.5 $|$ 4.3  \\
			&	2     &74.2 $|$ 98.6  &99.2 $|$ 98.4   &90.5 $|$ 92.8    &98.1 $|$ 99.8  \\
			&	3     &93.9 $|$ 97.0  &99.9 $|$ 99.2   &     -           &99.8 $|$ 99.9   \\
			&	4     &88.8 $|$ 97.6  &98.9 $|$ 100.0  &     -           &100.0 $|$ 100.0  \\
			&	5     &89.5 $|$ 94.3  &98.0 $|$ 100.0    &     -           &98.9 $|$ 99.7  \\
			&	6     &86.3 $|$ 95.9  &     -          &     -           &97.8 $|$ 98.5   \\
			\bottomrule  
			
		\end{tabular}
	\end{adjustbox}
\end{table*}

Based on the above experimental results, it can be seen that the performance of clean-label backdoor attack is worse than dirty-label backdoor attack. In deep learning, the model needs to establish a good mapping between the robust features of the sample itself and the label, which in turn enables accurate prediction. 
However, in the clean-label backdoor attack, the model needs to associate the trigger with its own label, hence the robust features of the sample itself can seriously interfere with the attack. Therefore, the performance of clean-label backdoor attacks is usually lower than that of modified label backdoor attacks.

\subsubsection*{\noindent \textbf{R4: How do hyperparameters affect the effectiveness of an attack?}}

\begin{figure*}[!ht]
	
	\centering
	\begin{minipage}[t]{0.98\linewidth} % 
		\centering
		\includegraphics[width=0.98\linewidth]{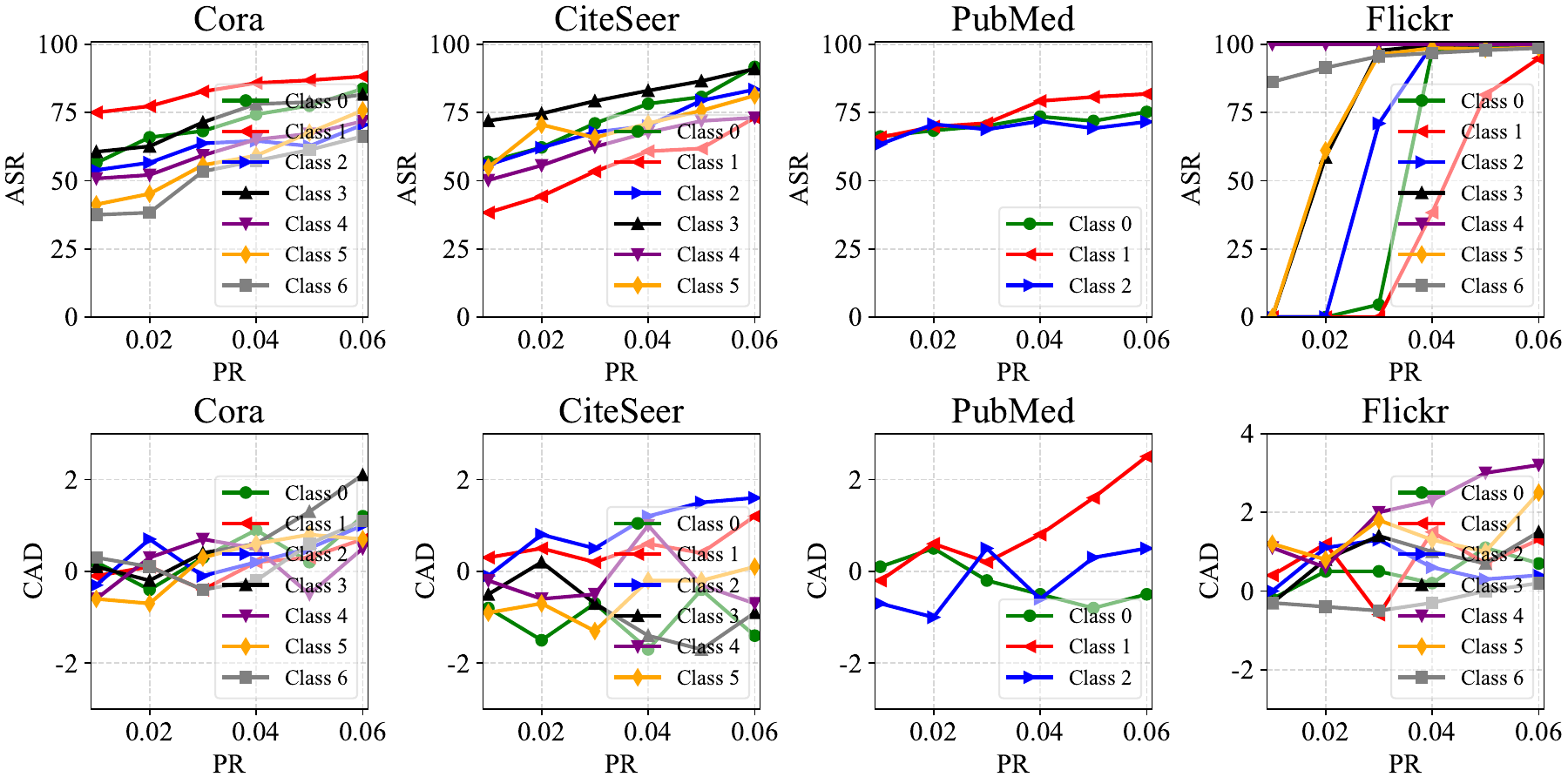}
	\end{minipage}
	\caption{The effect of PR on ASR and CAD. The model is GraphSAGE.  Trigger sizes in four datasets are 0.01, 0.005, 0.05, and 0.02, respectively.}
	\label{fig3}  % The label position need to be under the caption.
	
\end{figure*}

\noindent \textbf{The effect of poisoning rate}:
To investigate the effect of different poisoning rates (PR) on the effectiveness of the attacks, we set the PR to different values and conduct the experiments separately. The experimental results of different PRs on ASR and CAD are shown in Figure \ref{fig3}.

Overall, ASR shows a positive correlation with PR in different datasets. As PR increases, ASR shows an increasing trend, and this behavior is most obvious in Flickr.
For some classes in Flickr, the ASR shows explosive growth for each percentage point increase in PR in a specific range. In different classes, the sensitive area of ASR changes with PR is different. For example, in Class 4 and Class 6, a high attack success rate can be achieved when the poisoning rate is only 1\%, so the ASR changes less with increasing PR in this class.
However, for all other classes, the attacks are less effective when PR is less than 2\%, and ASR shows rapid growth as PR increases. However, the growth trend in the Cora, CiteSeer, and PubMed is relatively slow, especially in PubMed where the ASR increase trend is the slowest.
The reason for this is that in Cora, CiteSeer, and PubMed datasets, the number of nodes in each class is relatively balanced and the number of nodes in the training set is less, so the attack success rate is worse compared to Flickr. 
The primary reason for this phenomenon is that, in the Cora, CiteSeer, and PubMed, the number of nodes in each class is relatively balanced, and there are fewer nodes in the training set, resulting in lower attack success rates compared to Flickr. In contrast, the Flickr exhibits an uneven distribution of nodes among classes, with significant disparities. For instance, the node counts for Class 4 and Class 6 are 11,525 and 18,814, respectively, while the node counts for other classes are all below 5,000. This is a key factor explaining why Class 4 and Class 6 achieve higher attack success rates at lower poisoning rates.

Regarding CAD, with an increase in PR, CAD does not exhibit significant overall changes. Some classes show a slight upward trend in CAD, but others do not display noticeable variations. At PR=0.06, CAD across various classes is nearly all below 3\%. In other words, the insertion of triggers does not significantly impact the backdoor model's predictive accuracy on clean samples, indicating that CGBA achieves favorable attack performance.

\noindent \textbf{The effect of the degree of nodes on the experimental results}:
In a graph, nodes with large degree have more neighboring nodes, i.e., more similar nodes. We utilize the feature of that node as a trigger to increase the similarity of the nodes to improve the stealthiness of the trigger. Therefore, the above experimental results are the triggers selected in the node feature with the largest degree. To investigate whether the degree selection of a node affects the attack effect, we select the trigger in the node with the smallest degree for the second experiment. The experimental results are shown in Table \ref{tab9}.

\begin{table}[!htbp]
	\caption{Results on minimum degree nodes.  The model is GraphSAGE. The poisoning rate is set to 0.04. The trigger sizes are 0.01, 0.005, 0.05, and 0.02.  "-" indicates that the class does not exist in the dataset.} 
	\label{tab9}
	\centering
	\begin{adjustbox}{width=0.45\textwidth}  % Adjust table size
		\begin{tabular}{ccccc} 
			\toprule 
			\multirow{2}*{Class}  & \multicolumn{4}{c}{ASR(\%) $|$ CAD(\%)} \\
			\cmidrule(lr){2-5}
			
			&Cora         &CiteSeer     &PubMed          &Flickr            \\
			\midrule 
			
			0     &77.2 $|$ 0.2   &94.3 $|$ -0.9  & 53.9 $|$ -0.3  & 96.8 $|$ 2.1    \\
			1     &69.4 $|$ 1.3   &63.1 $|$ 0.8   & 72.9 $|$ 1.0   & 82.8 $|$ 0.6   \\
			2     &70.7 $|$ -0.4  &76.9 $|$ 0.3   & 76.3 $|$ -0.6  & 100.0 $|$ 1.1   \\
			3     &85.3 $|$ 1.1   &75.2 $|$ -1.3  &       -        & 88.9 $|$ 1.0    \\
			4     &51.8 $|$ 0.3   &63.8 $|$ 1.6   &       -        & 100.0 $|$ 4.3   \\
			5     &56.9 $|$ -0.8  &89.5 $|$ 0.2   &       -        & 26.6 $|$ 0.9   \\
			6     &99.7 $|$ -1.0  &    -          &       -        & 99.6 $|$ 0.2   \\
			
			\bottomrule  
		\end{tabular}
	\end{adjustbox}
\end{table}

When we select feature triggers from the minimum degree nodes, the attack success rate and all other metrics do not show a clear trend toward better or worse overall. In some classes of nodes, the attack success rate is better, but in some is lower. This illustrates that the degree of a node has little effect on the attack success rate, and it is more the features of the node that determine its label. 
Therefore, it is also necessary to explore the effect of different triggers on the experimental results.

\noindent \textbf{The effect of different feature triggers on the experimental results}:
In the previous experiments, triggers consisted of the maximum feature values selected from nodes belonging to the target class. To explore the influence of different feature triggers on the experimental results, we conduct experiments using the minimum feature triggers and random feature triggers. It is important to note that both random features and minimum features are selected from non-zero feature vectors.
Since the non-zero feature vectors of Cora and CiteSeer are sparse, only the PubMed and Flickr datasets are utilized for the experiments here.

\begin{figure*}[h]
	
	\centering
	\begin{minipage}[t]{0.9\linewidth} % 
		\centering
		\includegraphics[width=0.9\linewidth]{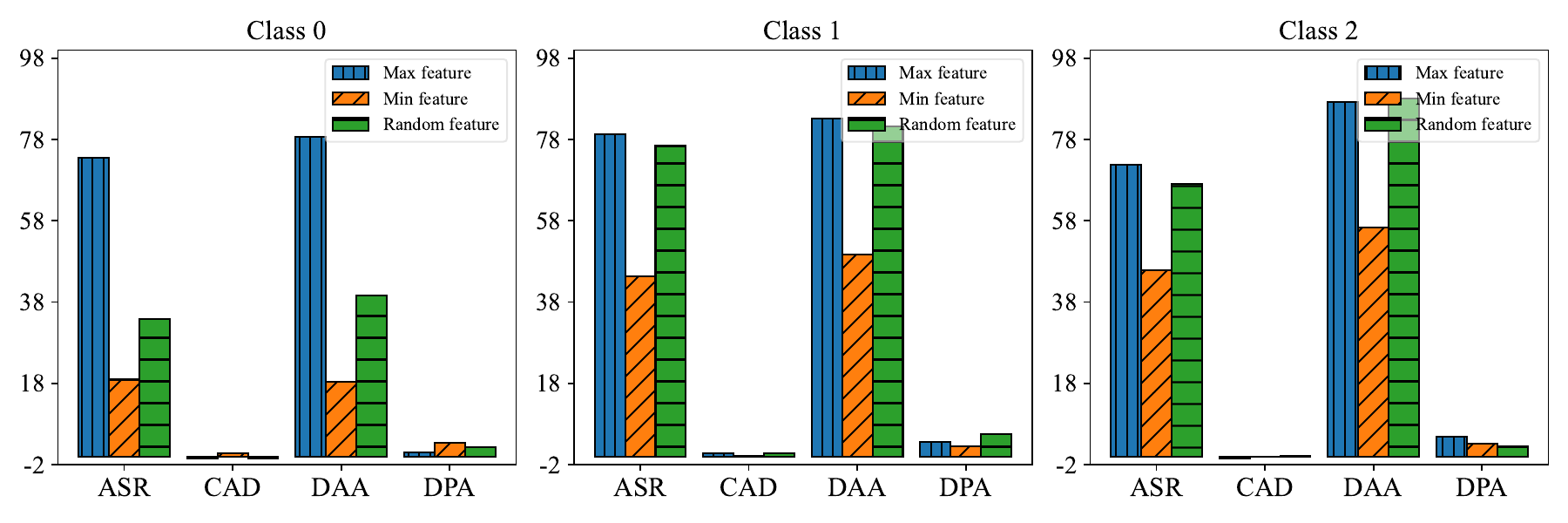}
	\end{minipage}
	\caption{Experimental results with different triggers in PubMed. The model is GraphSAGE. The PR is set to  0.04 and the trigger is selected among the nodes with the largest degree.}
	\label{fig4}
	
\end{figure*}

From the experimental results in Figure \ref{fig4}, it is evident that Max feature triggers achieve a higher attack success rate, while Min feature triggers exhibit the lowest success rate, with Random feature triggers falling in between. Regarding CAD and DPA, no significant variations are observed across different classes. In contrast, DAA displays a noticeable increase in numerical values compared to ASR, while still adhering to the same pattern, indicating that Max feature triggers deliver superior overall attack performance.

The experimental findings pertaining to the Flickr are provided in Table \ref{tab10}. Notably, Max feature triggers demonstrate a superior level of attack effectiveness compared to Min feature and Random feature triggers. In the case of Class 4, the attack performance of Random feature triggers surpasses that of Min feature triggers, while in other classes, Min feature and Random feature triggers exhibit a comparable level of attack effectiveness. It is noteworthy that both CAD and DPA metrics exhibit consistent behavior across various trigger types, showing no significant deviations or fluctuations.

\begin{table*}[!htbp]
	\caption{Experimental results with different triggers in Flickr. The PR is set to 0.04 and the trigger is selected among the nodes with the largest degree.} 
	\label{tab10}
	\centering
	\begin{adjustbox}{width=0.6\textwidth}  % Adjust table size
		\begin{tabular}{cccc} 
			\toprule 
			\multirow{2}*{Class}  & \multicolumn{3}{c}{ ASR(\%) $|$ CAD(\%) $|$ DAA(\%) $|$ DPA(\%)} \\
			\cmidrule(lr){2-4}
			
			&Max feature         &Min feature           &Random feature                   \\
			\midrule 
			
			0     & 97.5 $|$ 0.2 $|$ 95.5 $|$ 0.8   & 0.0 $|$ 0.4 $|$ 0.0 $|$ 1.1    &  0.0 $|$ 0.5 $|$ 0.0 $|$ -0.3 \\
			1     & 38.3 $|$ 1.5 $|$ 46.2 $|$ 0.1   & 0.0 $|$ -0.1 $|$ 0.0 $|$ 0.7   &  0.0 $|$ 1.1 $|$ 0.0 $|$ -0.1 \\
			2     & 99.8 $|$ 0.6 $|$ 99.7 $|$ 0.5   & 0.0 $|$ 0.3 $|$ 0.0 $|$ 0.8    &  0.0 $|$ 0.7 $|$ 0.0 $|$ 0.4 \\
			3     & 99.6 $|$ 1.0 $|$ 99.5 $|$ 0.7   & 0.0 $|$ 0.8 $|$ 0.0 $|$ 0.7    &  0.0 $|$ 0.2 $|$ 0.0 $|$ 0.3 \\
			4     & 100.0 $|$ 2.3 $|$ 99.9 $|$ -0.2 & 17.2 $|$ 0.5 $|$ 16.8 $|$ 0.6  &  100.0 $|$ 2.7 $|$ 99.8 $|$ -0.9 \\
			5     & 98.6 $|$ 1.3 $|$ 99.7 $|$ 1.3   & 0.0 $|$ 0.3 $|$ 0.0 $|$ 0.2    &  0.0 $|$ 0.6 $|$ 0.0 $|$ 0.4 \\
			6     & 96.8 $|$ -0.3 $|$ 97.2 $|$ 1.2  & 82.8 $|$ 0.3 $|$ 85.6 $|$ 0.5  &  83.7 $|$ 0.1 $|$ 88.2 $|$ 1.0 \\
			
			\bottomrule  
		\end{tabular}
	\end{adjustbox}
\end{table*}

\begin{figure*}[!h]
	
	\centering
	\begin{minipage}[t]{0.98\linewidth} % 
		\centering
		\includegraphics[width=0.9\linewidth]{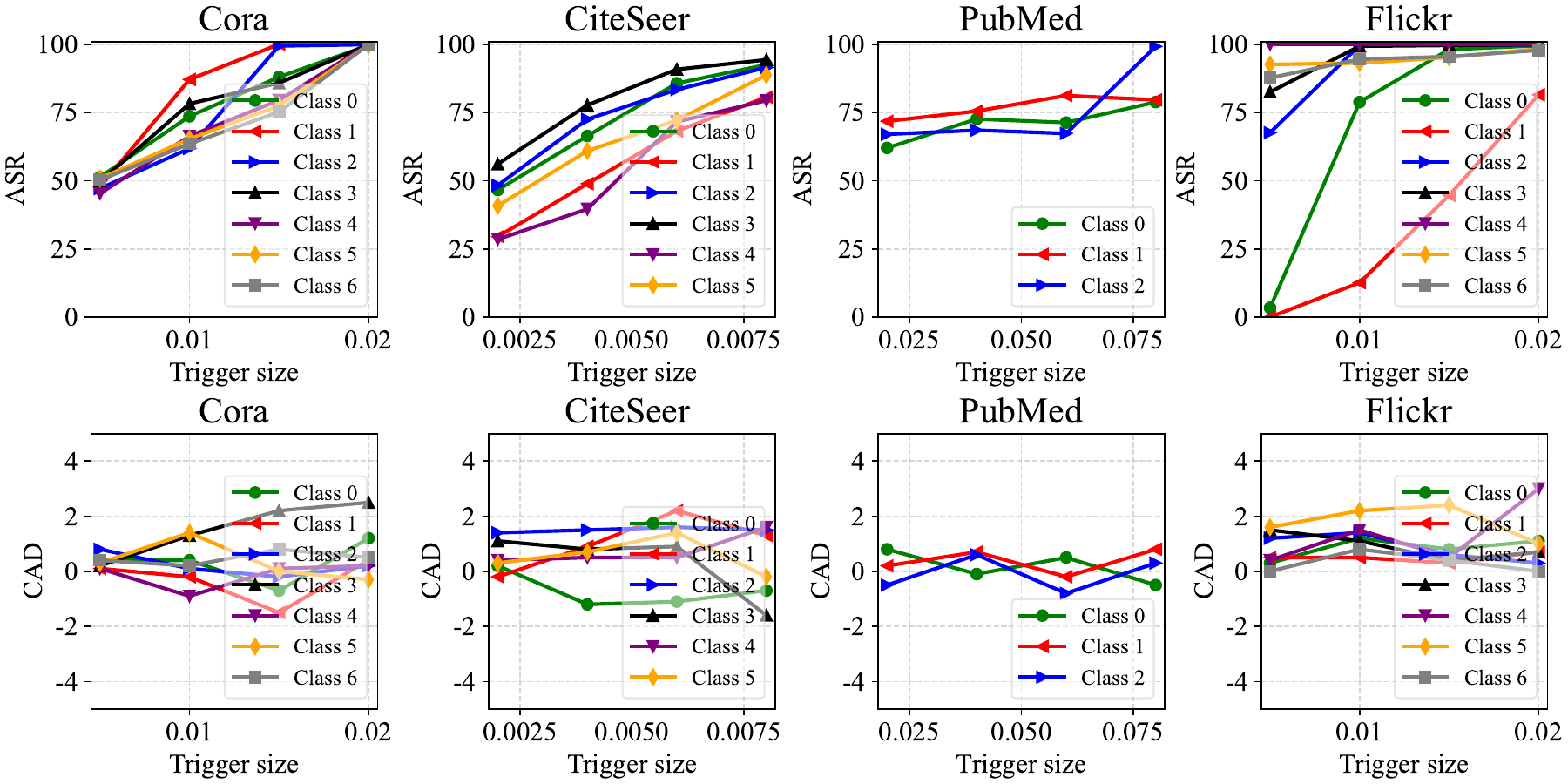}
		\label{fig7a}
	\end{minipage}
	\caption{The effect of trigger size on ASR and CAD. The PR is set to 0.05. The model is GraphSAGE. }
	\label{fig5}
	
\end{figure*}

\noindent \textbf{The effect of trigger size on experimental results}:
To investigate the effect of trigger size on the experimental results, the triggers are set to different sizes for the experiments in this paper. The experimental results of ASR and CAD for the four datasets are shown in Figure \ref{fig5}. Similar to PR, as the Trigger size increases, the attack success rate also shows a significant increase, but CAD still has no obvious trend.

\section{Conclusions}
\label{sec6}
In this paper, we propose CGBA, a clean-label backdoor attack scheme on node classification tasks. CGBA is a simple and effective attack scheme that does not modify the label of the nodes and the structure of the graph, and only injects subtle feature triggers to achieve high-performance attacks. To meet the similarity of neighboring nodes and avoid trigger edges being defended by cropping based on low-similarity edges, CGBA selects triggers in the node's own features to further enhance the concealment. Overall, our method is less perturbing to the training dataset and thus has higher backdoor concealment. Extensive experiments are conducted on four benchmark datasets to demonstrate the effectiveness of the CGBA attack. Since there is no backdoor defense method on the node classification task, we conduct defense experiments against CGBA using classical defense methods in graph adversarial attacks. According to the experimental results, we observe that the adversarial defense method based on node similarity not only fails to effectively defend against CGBA but also improves the success rate of CGBA attacks, which also reveals that other effective defense methods need to be explored against CGBA. Therefore, in our future work, we will focus on further exploring the backdoor vulnerability of GNNs and exploring more effective backdoor defense methods for GNNs.



\bibliographystyle{elsarticle-num} 
\bibliography{CGBA}

\begin{thebibliography}{10}
\expandafter\ifx\csname url\endcsname\relax
  \def\url#1{\texttt{#1}}\fi
\expandafter\ifx\csname urlprefix\endcsname\relax\def\urlprefix{URL }\fi
\expandafter\ifx\csname href\endcsname\relax
  \def\href#1#2{#2} \def\path#1{#1}\fi

\bibitem{1kumar2022influence}
S.~Kumar, A.~Mallik, A.~Khetarpal, B.~Panda, Influence maximization in social
  networks using graph embedding and graph neural network, Information Sciences
  607 (2022) 1617--1636, \url{https://doi.org/10.1016/j.ins.2022.06.075}.

\bibitem{2zhang2022improving}
Y.~Zhang, S.~Gao, J.~Pei, H.~Huang, Improving social network embedding via new
  second-order continuous graph neural networks, in: Proceedings of the 28th
  ACM SIGKDD Conference on Knowledge Discovery and Data Mining, 2022, p.
  2515–2523, \url{https://doi.org/10.1145/3534678.3539415}.

\bibitem{3gao2022graph}
C.~Gao, X.~Wang, X.~He, Y.~Li, Graph neural networks for recommender system,
  in: Proceedings of the Fifteenth ACM International Conference on Web Search
  and Data Mining, 2022, pp. 1623--1625,
  \url{https://doi.org/10.1145/3488560.3501396}.

\bibitem{4wu2022graph}
S.~Wu, F.~Sun, W.~Zhang, X.~Xie, B.~Cui, Graph neural networks in recommender
  systems: a survey, ACM Computing Surveys 55~(5) (2022) 1--37,
  \url{https://doi.org/10.1145/3535101}.

\bibitem{5hao2020asgn}
Z.~Hao, C.~Lu, Z.~Huang, H.~Wang, Z.~Hu, Q.~Liu, E.~Chen, C.~Lee, Asgn: An
  active semi-supervised graph neural network for molecular property
  prediction, in: Proceedings of the 26th ACM SIGKDD International Conference
  on Knowledge Discovery \& Data Mining, 2020, pp. 731--752,
  \url{https://doi.org/10.1145/3394486.3403117}.

\bibitem{6wang2022molecular}
Y.~Wang, J.~Wang, Z.~Cao, A.~Barati~Farimani, Molecular contrastive learning of
  representations via graph neural networks, Nature Machine Intelligence 4~(3)
  (2022) 279--287, \url{https://doi.org/10.1038/s42256-022-00447-x}.

\bibitem{7feng2022fiba}
Y.~Feng, B.~Ma, J.~Zhang, S.~Zhao, Y.~Xia, D.~Tao, Fiba: Frequency-injection
  based backdoor attack in medical image analysis, in: Proceedings of the
  IEEE/CVF Conference on Computer Vision and Pattern Recognition, 2022, pp.
  20876--20885, \url{https://doi.org/10.1109/CVPR52688.2022.02021 }.

\bibitem{8hua2023unambiguous}
G.~Hua, A.~B.~J. Teoh, Y.~Xiang, H.~Jiang, Unambiguous and high-fidelity
  backdoor watermarking for deep neural networks, IEEE Transactions on Neural
  Networks and Learning Systems\url{https://doi.org/10.1109/TNNLS.2023.3250210}
  (2023).

\bibitem{9gong2023kaleidoscope}
X.~Gong, Z.~Wang, Y.~Chen, M.~Xue, Q.~Wang, C.~Shen, Kaleidoscope: Physical
  backdoor attacks against deep neural networks with rgb filters, IEEE
  Transactions on Dependable and Secure
  Computing\url{https://doi.org/10.1109/TDSC.2023.3239225} (2023).

\bibitem{10li2022backdoors}
S.~Li, T.~Dong, B.~Z.~H. Zhao, M.~Xue, S.~Du, H.~Zhu, Backdoors against natural
  language processing: A review, IEEE Security \& Privacy 20~(05) (2022)
  50--59, \url{https://doi.org/10.1109/MSEC.2022.3181001}.

\bibitem{11pan2022hidden}
X.~Pan, M.~Zhang, B.~Sheng, J.~Zhu, M.~Yang, Hidden trigger backdoor attack on
  nlp models via linguistic style manipulation, in: 31st USENIX Security
  Symposium (USENIX Security 22), 2022, pp. 3611--3628,
  \url{https://dblp.org/rec/conf/uss/PanZSZ022}.

\bibitem{12xi2021graph}
Z.~Xi, R.~Pang, S.~Ji, T.~Wang, Graph backdoor, in: 30th USENIX Security
  Symposium (USENIX Security 21), 2021, pp. 1523--1540,
  \url{https://dblp.org/rec/conf/uss/XiPJ021}.

\bibitem{13zheng2023link}
H.~Zheng, H.~Xiong, H.~Ma, G.~Huang, J.~Chen, Link-backdoor: Backdoor attack on
  link prediction via node injection, IEEE Transactions on Computational Social
  Systems\url{https://doi.org/10.1109/TCSS.2023.3260833} (2023).

\bibitem{14dai2023semantic}
J.~Dai, Z.~Xiong, A semantic backdoor attack against graph convolutional
  networks, arXiv preprint
  arXiv:2302.14353\url{https://doi.org/10.48550/arXiv.2302.14353} (2023).

\bibitem{15zhang2021backdoor}
Z.~Zhang, J.~Jia, B.~Wang, N.~Z. Gong, Backdoor attacks to graph neural
  networks, in: Proceedings of the 26th ACM Symposium on Access Control Models
  and Technologies, 2021, pp. 15--26,
  \url{https://doi.org/10.1145/3450569.3463560}.

\bibitem{16zheng2023motif}
H.~Zheng, H.~Xiong, J.~Chen, H.~Ma, G.~Huang, Motif-backdoor: Rethinking the
  backdoor attack on graph neural networks via motifs, IEEE Transactions on
  Computational Social Systems\url{https://doi.org/10.1109/TCSS.2023.3267094}
  (2023).

\bibitem{17alrahis2023poisonedgnn}
L.~Alrahis, S.~Patnaik, M.~A. Hanif, M.~Shafique, O.~Sinanoglu, Poisonedgnn:
  Backdoor attack on graph neural networks-based hardware security systems,
  IEEE Transactions on Computers\url{https://doi.org/10.1109/TC.2023.3271126}
  (2023).

\bibitem{18chen2022neighboring}
L.~Chen, Q.~Peng, J.~Li, Y.~Liu, J.~Chen, Y.~Li, Z.~Zheng, Neighboring backdoor
  attacks on graph convolutional network, arXiv preprint
  arXiv:2201.06202\url{https://doi.org/10.48550/arXiv.2201.06202 } (2022).

\bibitem{34zhang2023graph}
H.~Zhang, J.~Chen, L.~Lin, J.~Jia, D.~Wu, Graph contrastive backdoor
  attacks\url{https://dblp.org/rec/conf/icml/ZhangCLJW23} (2023).

\bibitem{19chen2023feature}
Y.~Chen, Z.~Ye, H.~Zhao, Y.~Wang, et~al., Feature-based graph backdoor attack
  in the node classification task, International Journal of Intelligent Systems
  2023, \url{https://doi.org/10.1155/2023/5418398 } (2023).

\bibitem{20dai2023unnoticeable}
E.~Dai, M.~Lin, X.~Zhang, S.~Wang, Unnoticeable backdoor attacks on graph
  neural networks, in: Proceedings of the ACM Web Conference 2023, 2023, pp.
  2263--2273, \url{https://doi.org/10.1145/3543507.3583392}.

\bibitem{22li2022reliable}
K.~Li, Y.~Liu, X.~Ao, J.~Chi, J.~Feng, H.~Yang, Q.~He, Reliable representations
  make a stronger defender: Unsupervised structure refinement for robust gnn,
  in: Proceedings of the 28th ACM SIGKDD Conference on Knowledge Discovery and
  Data Mining, 2022, pp. 925--935,
  \url{https://doi.org/10.1145/3534678.3539484}.

\bibitem{31xu2021explainability}
J.~Xu, M.~Xue, S.~Picek, Explainability-based backdoor attacks against graph
  neural networks, in: Proceedings of the 3rd ACM Workshop on Wireless Security
  and Machine Learning, 2021, pp. 31--36,
  \url{https://doi.org/10.1145/3468218.3469046}.

\bibitem{32jiang2022defending}
B.~Jiang, Z.~Li, Defending against backdoor attack on graph nerual network by
  explainability, arXiv preprint
  arXiv:2209.02902\url{https://doi.org/10.48550/arXiv.2209.02902} (2022).

\bibitem{33chen2023contrast}
j.~Chen, H.~Xiong, H.~Ma, Y.~Zheng, Clb-defense: based on contrastive learning
  defense for graph neural network against backdoor attack, Journal on
  Communications 44~(4) (2023) 154--166,
  \url{https://www.infocomm-journal.com/txxb/EN/Y2023/V44/I4/154}.

\bibitem{26turner2018clean}
A.~Turner, D.~Tsipras, A.~Madry, Clean-label backdoor
  attacks\url{https://openreview.net/forum?id=HJg6e2CcK7} (2018).

\bibitem{35ning2021invisible}
R.~Ning, J.~Li, C.~Xin, H.~Wu, Invisible poison: A blackbox clean label
  backdoor attack to deep neural networks, in: IEEE INFOCOM 2021-IEEE
  Conference on Computer Communications, IEEE, 2021, pp. 1--10,
  \url{https://doi.org/10.1109/INFOCOM42981.2021.9488902}.

\bibitem{36gao2023not}
Y.~Gao, Y.~Li, L.~Zhu, D.~Wu, Y.~Jiang, S.-T. Xia, Not all samples are born
  equal: Towards effective clean-label backdoor attacks, Pattern Recognition
  139 (2023) 109512, \url{https://doi.org/10.1016/j.patcog.2023.109512}.

\bibitem{25xu2022poster}
J.~Xu, S.~Picek, Poster: Clean-label backdoor attack on graph neural networks,
  in: Proceedings of the 2022 ACM SIGSAC Conference on Computer and
  Communications Security, 2022, pp. 3491--3493,
  \url{https://doi.org/10.1145/3548606.3563531}.

\bibitem{23kipf2016semi}
T.~N. Kipf, M.~Welling, Semi-supervised classification with graph convolutional
  networks, arXiv preprint
  arXiv:1609.02907\url{https://doi.org/10.48550/arXiv.1609.02907} (2016).

\bibitem{24velivckovic2017graph}
P.~Veli{\v{c}}kovi{\'c}, G.~Cucurull, A.~Casanova, A.~Romero, P.~Lio,
  Y.~Bengio, Graph attention networks, arXiv preprint
  arXiv:1710.10903\url{https://doi.org/10.48550/arXiv.1710.10903} (2017).

\bibitem{27yang2016revisiting}
Z.~Yang, W.~Cohen, R.~Salakhudinov, Revisiting semi-supervised learning with
  graph embeddings, in: International conference on machine learning, 2016, pp.
  40--48, \url{https://dblp.org/rec/conf/icml/YangCS16}.

\bibitem{28zeng2019graphsaint}
H.~Zeng, H.~Zhou, A.~Srivastava, R.~Kannan, V.~Prasanna, Graphsaint: Graph
  sampling based inductive learning method, arXiv preprint
  arXiv:1907.04931\url{https://doi.org/10.48550/arXiv.1907.04931} (2019).

\bibitem{29hamilton2017inductive}
W.~Hamilton, Z.~Ying, J.~Leskovec, Inductive representation learning on large
  graphs, Advances in neural information processing systems 30,
  \url{https://dblp.org/rec/conf/nips/HamiltonYL17 } (2017).

\bibitem{37Chebnet2016}
M.~Defferrard, X.~Bresson, P.~Vandergheynst, Convolutional neural networks on
  graphs with fast localized spectral filtering, Advances in neural information
  processing systems 29,
  \url{https://proceedings.neurips.cc/paper/2016/hash/04df4d434d481c5bb723be1b6df1ee65-Abstract.html}
  (2016).

\bibitem{38ARMA2021}
F.~M. Bianchi, D.~Grattarola, L.~Livi, C.~Alippi, Graph neural networks with
  convolutional arma filters, IEEE transactions on pattern analysis and machine
  intelligence 44~(7) (2021) 3496--3507,
  \url{https://doi.org/10.1109/TPAMI.2021.3054830}.

\bibitem{30kingma2014adam}
D.~P. Kingma, J.~Ba, Adam: A method for stochastic optimization, arXiv preprint
  arXiv:1412.6980\url{https://doi.org/10.48550/arXiv.1412.6980} (2014).

\end{thebibliography}


\end{document}